# Properties of the $a_0(980)$ Meson

S. Teige[1], G. S. Adams[6], T. Adams[2], E. V. Anoshina[8], Z. Bar-Yam[4], J. M. Bishop[2], V. A. Bodyagin[8], B. B. Brabson[1], D. S. Brown[5], N. M. Cason[2], S. U. Chung[3], R. R. Crittenden[1], J. P. Cummings[4], S. Denisov[7], J. P. Dowd[4], A. Dushkin[7], A. R. Dzierba[1], P. Eugenio[4], A. M. Gribushin[8], J. Gunter[1], R. W. Hackenburg[3], M. Hayek[4], W. Kern[4], E. King[4], V. Kochetkov[7], O. L. Kodolova[8], V. L. Korotkikh[8], M. A. Kostin[8], R. Lindenbusch[1], J. M. LoSecco[2], J. J. Manak[2], J. Napolitano[6], M. Nozar[6], C. Olchanski[3], A. I. Ostrovidov[8], T. K. Pedlar[5], A. S. Proskuryakov[8], D. R. Rust[1], A. H. Sanjari[2], L. I. Sarycheva[8], E. Scott[1], K. K. Seth[5], I. Shein[7], W. D. Shephard[2], N. B. Sinev[8], J. A. Smith[6], P. T. Smith[1], A. Soldatov[7], D. L. Stienike[2], T. Sulanke[1], S. A. Taegar[2], D. R. Thompson[2], I. N. Vardanyan[8], D. P. Weygand[3], H. J. Willutzki[3], J. Wise[5], M. Witkowski[6] A. A. Yershov[8], D. Zhao[5],

[1] *Department of Physics, Indiana University,Bloomington IN 47405, USA*

[2] *Department of Physics, University of Notre Dame, Notre Dame IN 46556, USA*

[3] *Department of Physics, Brookhaven National Laboratory,Upton, L.I., NY 11973*

[4] *Department of Physics, University of Massachusetts Dartmouth,North Dartmouth, MA 02747,USA*

[5] *Department of Physics, Northwestern University,Evanston, IL 60208, USA*

[6] *Department of Physics, Ressselaer Polytechnic Institute, Troy NY,USA* [7] *Institute for High Energy Physics,Protovino, Russian Federation*

[8] *Institute for Nuclear Physics, Moscow State University,Moscow, Russian Federation*

The mass and width of the $a_0(980)$ have been independently determined from a nearly background free data sample and the coupling constants to the $\eta\pi$ and $K\overline{K}$ modes have been determined.

## 1 Introduction

This paper presents the results of an analysis of the reaction $\pi^- p \rightarrow \eta\pi^+\pi^- n$ at 18.3 GeV/$c^2$. The data were taken during the 1994 running period of the Brookhaven National Laboratory (BNL) AGS by the E852 collaboration. E852, A Search for Mesons with Exotic Quantum Numbers, was performed at the BNL Multiparticle Spectrometer augmented by nearly hermetic photon detection and vetoing system, an instrumented target region and additional downstream tracking.

The photon detection system consisted of a large, segmented lead glass calorimeter (LGD) and and the photon veto system was comprised of a lead scintillator sandwich photon veto counter (DEA) and a segmented Cesium Iodide detector. [1] The performance characteristics of prototypes of the LGD have been described previously [2] and the detector used here behaved similarly.

The DEA was a “picture frame” type counter with an aperture chosen so that large angle photons that would miss the LGD would be intercepted by the DEA. This device was used as a veto counter in the trigger for the data analyzed here. Additionally, a 4-layer cylindrical drift tube detector [3] surrounded the target. These systems, taken together, allowed us to select well contained events with a neutron recoil.

The LGD was also used as a trigger device in conjunction with a custom built trigger processor. This processor [4] could determine the total energy deposited in the LGD and the effective mass of the photon system. The data taken for this analysis required a photon system with an effective mass greater than the mass of the $\pi^0$ .

Figure 1.1 shows the observed $\eta\pi^+\pi^-$ effective mass distribution observed. Clear structure associated with the $\eta'$ is observed. Additionally, a structure at 1285 MeV/c$^2$ is seen. A Dalitz plot analysis of the 3-body effective mass region centered on 1285 MeV/c$^2$ has shown that this structure is associated with a state (or states) that decay into $a_0\pi$

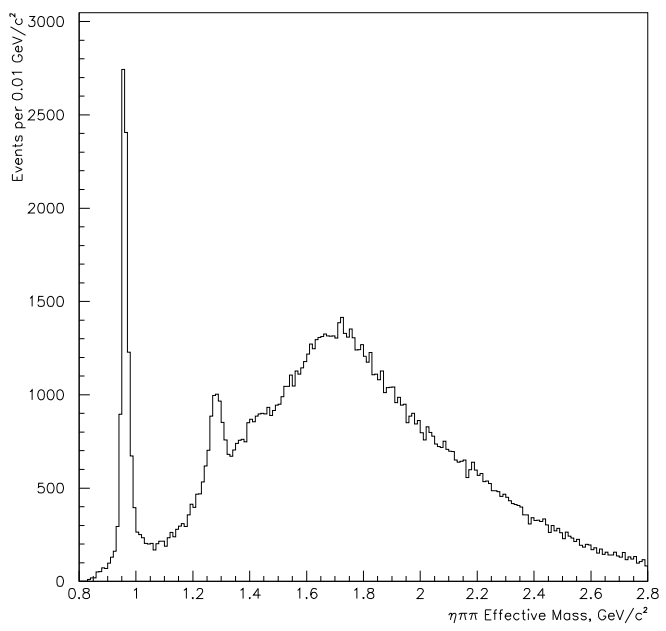

**Figure 1.1** The $\eta\pi^+\pi^-$ effective mass distribution.

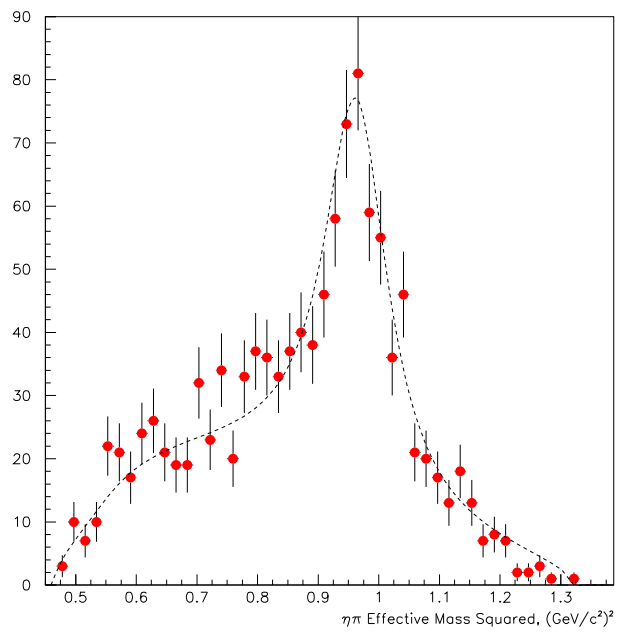

**Figure 1.2** The $\eta\pi^+$ effective mass distribution after a selection on the three body effective mass.

Figure 1.2 shows the $\eta\pi$ effective mass distribution after a selection on the three body effective mass. It is clear that this selection produces a relatively pure sample of $a_0$ decays. The mass and width of the $a_0$ is determined two ways, a fit to a Breit-Wigner plus polynomial background (to allow comparison with previous results [5]) and a fit to the form of Flattè [6].

The simple Breit-Wigner fit (shown in figure 1.2) gives a mass of $978 \pm 3$

MeV/$c^2$ and a width of $65 \pm 9$ MeV/$c^2$ . The $\eta\pi$ lineshape as given by the Flattè form is the square of the amplitude A where

$$A = \frac{\sqrt{\Gamma_0 \Gamma_{\eta\pi}} \mathrm{M_r}}{\mathrm{M_r}^2 - \mathrm{M}^2 - i\mathrm{M_r}(\Gamma_{\eta\pi} + \Gamma_{\overline{K}K})}, \quad (1)$$

$$\Gamma_{\eta\pi} = \mathrm{g}_{\eta\pi} q \quad (2)$$

and

$$\Gamma_{\overline{K}K} = i\mathrm{g}_{\overline{\mathrm{K}}\mathrm{K}} \sqrt{\mathrm{M_K}^2 - (\mathrm{M}/2)^2}. \quad (3)$$

$\mathrm{g}_{\eta\pi}$ and $\mathrm{g}_{\overline{\mathrm{K}}\mathrm{K}}$ are coupling constants determined by a fit to be $\mathrm{g}_{\eta\pi} = 0.40 \pm 0.02$ and $\mathrm{g}_{\overline{\mathrm{K}}\mathrm{K}} = 0.29 \pm 0.03$. The intensity implied by eqn. 1 (and a similar expression for the line shape for the $\overline{K}K$ mode) can be integrated to give the branching ratio

$$\frac{\Gamma(\mathrm{a}_0 \to \overline{\mathrm{K}}\mathrm{K})}{\Gamma(\mathrm{total})} = (14 \pm 2)\% \quad (4)$$

The integration has been carried out from $\eta\pi$ threshold to 1500 MeV/$c^2$ .

To conclude, the mass and width of the $\mathrm{a}_0(980)$ have been independently redetermined from a nearly background free data sample and the coupling constants for its two dominant decay modes have been determined.